\documentclass{nature}
\usepackage{graphicx}
\usepackage{verbatim}
\usepackage{color}
\usepackage[none]{hyphenat}
\usepackage{amsmath}
\usepackage[table,xcdraw]{xcolor}
\usepackage{stackengine}

\newcommand{\fourBullets}{\fbox{\Shortstack{$\bullet$ $\bullet$}\Shortstack{$\bullet$ $\bullet$}}}
\newcommand{\twoBullets}{\fbox{\Shortstack{$\bullet$ $\bullet$}}}

\title{Quantum dissection of a covalent bond}
\author{Norm M. Tubman${}^1$ and D.~ChangMo~Yang${}^1$}

\begin{document}
\maketitle
\begin{affiliations}
\item University of Illinois at Urbana-Champaign, Urbana, Illinois, USA
\end{affiliations}

\begin{abstract}

We propose that spatial density matrices, which are singularly important  in the
study of quantum entanglement\cite{review-1}, encode the electronic
fluctuations and correlations responsible for covalent bonding. From these
density matrices, we develop tools that allow us to analyse how 
many body wave functions can be broken up into real space pieces.  We apply
these tools to the first row dimers, and in particular, we address the
conflicting evidence in the literature about the presence of an inverted fourth
bond and anti-ferromagnetic correlations in the $\text{C}_2$
molecule\cite{cbond-1,trialogue,diss1,xu:2014}.  Our results show that many body
effects enhance anti-ferromagnetic fluctuations but are not related to the
formation of an inverted fourth bond.  We identify two inverted bonds in the
$\text{C}_2$ molecule and establish their correspondence to the bonds in the
$\text{Be}_2$ molecule. Additionally, we provide a new interpretation of the
Mayer index\cite{mayer-acc}, introduce partial bonds to fix deficiencies in
molecular orbital theory, and prove the Hartree-Fock wave function for C$_{2}$
is not a triple bond.  Our results suggest that entanglement-based methods can
lead to a more realistic treatment of molecular and extended systems than
possible before.  

\end{abstract}

Comparing the results from different theories of bonding is difficult, as these
theories often access very different properties with widely varying constraints
on the wave function.
Common types of bonding analysis include the use of molecular orbital~(MO)
theory to calculate bonding/anti-bonding pairs\cite{motheory}, density-matrix
renormalization group~(DMRG) to calculate mutual information of molecular
orbitals\cite{garnet2013}, and valence bond~(VB) theory to find the dominant
spin configurations\cite{cbond-1,xu:2014}.
These tools are
complemented by several different bond indices\cite{wiberg:1968,mayer-acc},
and theories based on single-particle properties, such as Bader
analysis\cite{bader:book}. 

Arguably, the real space degrees of freedom of a many body wave function are
the main quantities of interest that bonding analysis techniques are trying to
access.  VB theory comes close in this regard, as the optimised orbitals are
partially localised. However, this localisation requires a highly constrained
wave function 
 (see Table~\ref{tab:ene} for C$_{2}$ energies).  Other techniques
attempt to capture various properties related to bonding, but there has been a
lack of consensus of which provide the best description.  For instance,
new theoretical studies of the $\text{C}_{2}$ dimer have challenged conventional
descriptions of covalent bonding.  As an example of the divergent views on the
$\text{C}_2$ molecule\cite{cbond-1,trialogue,diss1,xu:2014}, a recent VB study
suggested an inverted fourth bond caused by many body
correlations\cite{cbond-1}, while a follow-up study showed the molecule is anti-ferromagnetically coupled~\cite{xu:2014}.  Recent DMRG simulations do not
have access to the proper degrees of freedom to settle the issue\cite{epfl},
and it is not clear that other techniques can be used to probe such features.
Another problem in describing covalent bonds is that many of these theories can
only be applied to very specific wave functions.  Explicit many body theories
such as mutual information studies in DMRG and spin configuration identification
in VB theory, cannot be applied to general wave functions optimized with variational Monte
Carlo~(VMC) or Hartree-Fock~(HF) wave functions.
This makes analysing even HF wave functions challenging.  Although there are
many systems for which MO theory provides a useful description of bonding
properties, the $\text{C}_{2}$ molecule is not one of them.
 
Real space density matrices can be used as a general tool to analyse bonds.
Their usefulness in determining quantum entanglement properties has garnered
widespread recognition in the past
decade\cite{review-1,tubman2,1dint-1,renyi-1,widom-1}, and the idea
itself goes back even further\cite{white:1992,dmrg}.
These density matrices
have been instrumental in recent breakthroughs with regards to applications in
string theory\cite{ryu:2006}, black holes\cite{review-1}, and topological phase
transitions\cite{jiang:2012}.  These same density matrices also happen to encode
how molecules form covalent bonds.
This analysis involves dividing a quantum system into two spatial regions, A and
B, which can be an arbitrary geometric bi-partition of space.  The degrees of
freedom in region~B are integrated out, leaving a density matrix that only
exists in region~A.  
This is written as $\hat{\rho}_{A} = \text{Tr}_{B}\{\hat{\rho}_{AB}\}$,
where $\text{Tr}_{B}$ is a trace over the degrees of freedom in region~B and
$\hat{\rho}_{AB}$ is the full density matrix of a system.  The density matrix
$\hat{\rho}_{B}$ can likewise be defined with a trace over region~A.  For the
diatomic molecules in this work, we will take regions~A and B to be the half
spaces on either side of the perpendicular bisector of the molecular axis.

A Schmidt decomposition of a wave function\cite{dmrg,peschel-func} can be used
to relate the spatial density matrices to the full many-body wave function as
\begin{equation}
  |\Psi\rangle = \sum_{i} s_{i}|A_{i}\rangle|B_{i}\rangle \; .
  \label{eq:fullent}
\end{equation}
Here, $|A_{i}\rangle$ and $|B_{i}\rangle$ are eigenvectors of $\hat{\rho}_{A}$
and $\hat{\rho}_{B}$ respectively, and $|s_{i}|^{2}$ are the eigenvalues.  The
eigenvalues of $\hat{\rho}_{A}$ and $\hat{\rho}_{B}$ are in one-to-one
correspondence and are equal, regardless of the partition of space.  
For the symmetric half-space partition of a homonuclear dimer, these density
matrices have mirror-symmetric eigenvectors.  

\begin{figure}
  \centering
  \includegraphics[height=0.65\textheight]{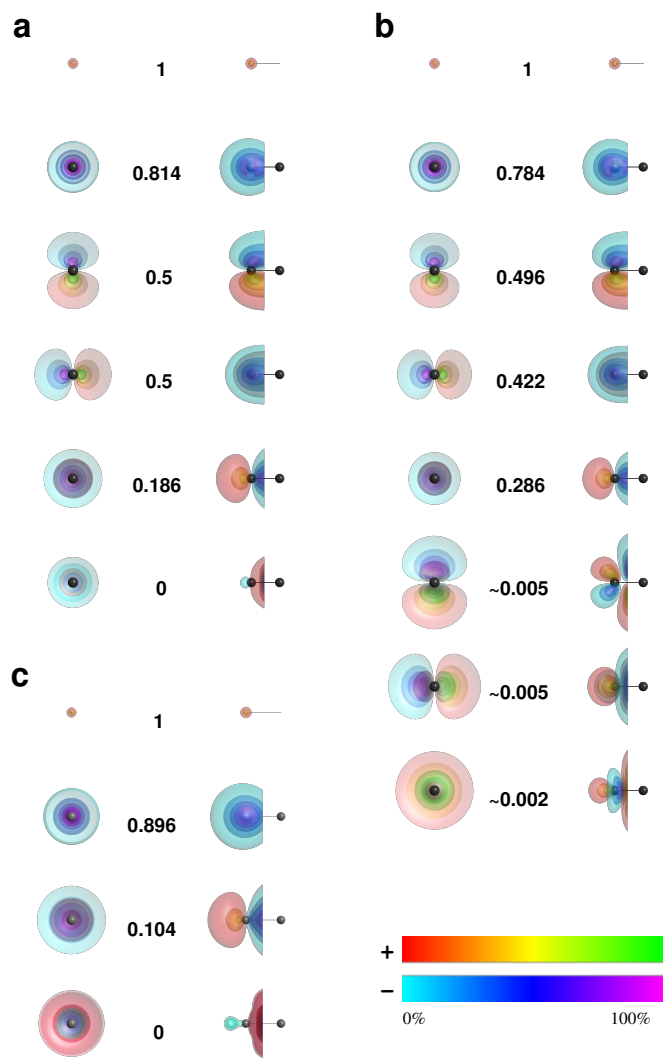} 
  \caption{Isosurfaces of entanglement natural orbitals for (a)~HF $\text{C}_2$,
(b)~Full valence CAS $\text{C}_2$, and (c)~HF $\text{Be}_2$.  The numerical
values are the $\lambda$ values, as described in the text, for the HF wave
functions, and they are the eigenvalues of the entanglement natural orbitals for
the full valence CAS wave function.  The plots show a view of the orbital along
the bonding axis (left), and a side view of the bond (right).  The isosurfaces
were chosen by the amount of integrated density less than the isosurface value,
which is indicated by the colour scale. There are actually 12 orbitals for the HF waveunfctions of 
$\text{C}_2$ and 8 orbitals for $\text{Be}_2$, but only half are plotted as
they are spin degenerate.   The orbitals corresponding to $\lambda=$ 0.814, 0.186
for HF $\text{C}_2$, and $\lambda=$ 0.896, 0.104 for HF $\text{Be}_2$ are inverted bonds.
Corresponding orbitals of nearly identical character can be seen in the full valence CAS $\text{C}_2$ orbitals with eigenvalues 0.784 and 0.286.
\label{fig:halforbs}}
\end{figure}

For a single-determinant wave function,
Eq.~\ref{eq:fullent} reduces to a more compact form 
\begin{equation}
  |\Psi\rangle  = \prod_{i=1}^{N} 
  \left(\sqrt{\lambda_{i}}c^{\dag}_{i,A}
  + \sqrt{1-\lambda_{i}}c^{\dag}_{i,B} \right)|0\rangle
\label{eq:hfent}
\end{equation}
with $2^N$ terms, where $N$ is the number of electrons\cite{peschel-func}.
In this equation, the values of $\lambda$ are in the range $0 \leq \lambda \leq
1$ and determine the probability distribution of electrons between regions~A
and~B.  The operator $c^{\dag}_{i,A}$ creates a single-particle orbital that has
support only in region~A, and $c^{\dag}_{i,B}$ creates a single-particle orbital
that has support only in region~B.  Quite simply, this equation says that an
electron $i$ is found with probability $\lambda_{i}$ in region~A and probability
$1-\lambda_{i}$ in region~B.  The case $\lambda = 1$ corresponds to an electron
that is completely localised in region~A and $\lambda = 0$ corresponds to an
electron completely localised in region~B.  The case of $\lambda = 1/2$
corresponds to an electron with equal probability to be in either regions~A
and~B.  Examples of region~A orbitals are plotted for $\text{C}_{2}$ in
Fig.~\ref{fig:halforbs}.

By introducing a single-determinant wave function in the form of
Eq.~\ref{eq:hfent}, we have provided a new theory for analysing covalent bonds.
We study its properties by first demonstrating its relationship with MO theory.
In MO theory, the occupation of a bonding orbital by two electrons is implicitly
interpreted as them being equally shared by the two atoms.
In Eq.~\ref{eq:hfent}, this corresponds to two electrons with $\lambda=1/2$
which can be expanded as
\begin{equation}
|\Psi\rangle = \frac{1}{{2}}|\phi_{\uparrow 1,A}\phi_{\downarrow 1,A}\rangle
+ \frac{1}{{2}}|\phi_{\uparrow 1,B}\phi_{\downarrow 1,A}\rangle
+ \frac{1}{{2}}|\phi_{\uparrow 1,A}\phi_{\downarrow 1,B}\rangle
+ \frac{1}{{2}}|\phi_{\uparrow 1,B}\phi_{\downarrow 1,B}\rangle \; .
\label{singlebond:eqn}
\end{equation} 
These four terms have a simple interpretation:  25\% chance both electrons are
localised in region~A, 25\% chance a spin-up electron is localised in region~A,
25\% chance a spin-down electron is localised in region~A, and 25\% chance that
neither are localised in region~A.  

When four electrons occupy a pair of bonding and anti-bonding orbitals, they are interpreted as not contributing to the bonding in MO theory.
The corresponding picture given by Eq.~\ref{eq:hfent} is a group of four
$\lambda$ values $(0,0,1,1)$, representing two electrons in region~A and two
electrons in region~B.  For this case, Eq.~\ref{eq:hfent} reduces to a single term.
Although $\lambda$ can take values from a continuous range, MO theory only
attempts to describe $\lambda = 0, 1/2,~\text{or}~ 1$.
For the molecules considered in this work, the ground state wave functions are  
mirror-symmetric with respect to the half space.  This leads to a important
constraint on the values of $\lambda$.  Whenever a spin-up electron is partially
localised in region~A, another spin-up electron must be localised in region~B to
maintain mirror symmetry of the charge between regions~A and B.  To maintain
spin symmetry, two more spin-down electrons must be introduced.
This couples electrons into groups of four such that
$\lambda_{1,2,3,4} = (\lambda, \lambda, 1-\lambda, 1-\lambda)$.
The orbitals for the first two electrons are the same except they have opposite
spin quantum numbers, and the orbitals for the third and fourth electrons are
also the same but with opposite spin.  
For these symmetric diatomic molecules, all electrons must come in groups
of four, with only one exception, $\lambda=1/2$.  In this case, it is possible
to maintain the proper space and spin symmetry with only two electrons.  This is
because mirror-symmetric orbitals can occur only for $\lambda = 1/2$.  

The values of $\lambda$ for HF wave functions are listed in
Table~\ref{tab:lambda}.  We see that the perfectly localised groups of four
$(0,0,1,1)$ and delocalised groups of two $(1/2,1/2)$ are the only grouping of
electrons needed to describe most of the molecules.  However, in the case of
$\text{C}_{2}$ and $\text{Be}_{2}$, there are groupings of four electrons that
are not perfectly localised and are not described by MO theory.  
For $\text{C}_{2}$, there has always been some uncertainty with in how to
describe the valence electrons. 
Various bond indices\cite{mayer-acc} yield values between 3 and 4, which
suggests the possibility of three strong bonds and one weak bond\cite{cbond-1}.
This is shown not to be the case in Table~\ref{tab:lambda}, where four of the
electrons have 50\% probability of being on either atom (corresponding to
orbitals with $\pi$-like symmetry, as shown in Fig.~\ref{fig:halforbs}), while
each of the remaining four electrons correspond to 81\% partial localisation.
An interesting aspect of this localisation is that the orbital amplitudes for
the paired orbitals $\lambda = 0.81$ and $\lambda = 0.19$ are not symmetric.
These orbitals are shown in Fig.~\ref{fig:halforbs} and have similar properties
as what has been previously interpreted as an inverted bond.  For $\lambda =
0.81$, the half-space orbital has rotational symmetry around the bonding axis,
and the corresponding orbital with $\lambda = 0.19$ also has this symmetry but
changes sign along the bonding access and has charge accumulation away from the
bond.  A key difference from our result and the previously described inverted
bond\cite{cbond-1}, is that we predict two of them from the HF wave function.
Our results definitively show that a multi-determinant wave function is not necessary
to capture the inverted bond --- it has always been present in the HF wave function. 

The result from our symmetry analysis explicitly prevents the HF wave function
of the $\text{C}_{2}$ molecule from being a triple bond.   For a triple bond to
occur in C$_{2}$, six of the valence electrons must be grouped differently from
the remaining two localised valence electrons.  But
symmetries force localised electrons to come in groups of four.  There are only
three possible groupings for eight electrons:
a quadruple bond $(\frac{1}{2}, \frac{1}{2}),(\frac{1}{2},\frac{1}{2}),(\frac{1}{2},\frac{1}{2}),(\frac{1}{2},\frac{1}{2})$;
a double bond with two partial bonds
$(\frac{1}{2}, \frac{1}{2})$, $(\frac{1}{2}, \frac{1}{2})$, $(\lambda, \lambda, 1-\lambda, 1-\lambda)$;
four partial bonds $(\lambda, \lambda, 1-\lambda, 1-\lambda)$, $(\lambda', \lambda', 1-\lambda', 1-\lambda')$.
Therefore it is impossible for eight valence electrons to form a triple
bond in HF theory without breaking spin and space symmetries.  

\begin{figure}
  \centering
  \includegraphics[width=0.55\textwidth]{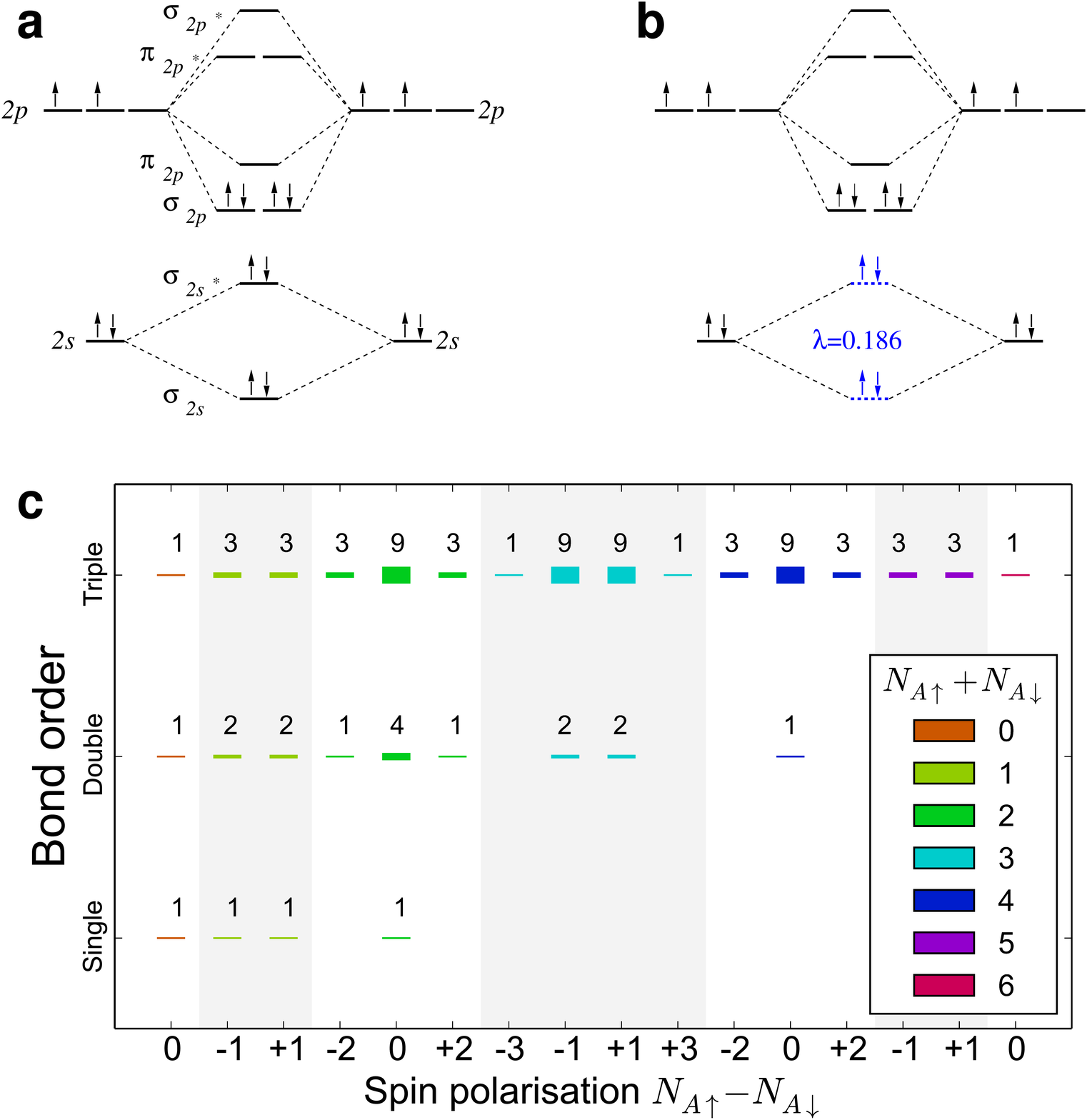}
  \caption{(a)~Orbital hybridisation of the valence electrons in $\text{C}_2$
according to MO theory.  The filling of three bonding and one anti-bonding
orbital is a double bond in MO theory.
(b)~Modification of the hybridisation diagram with entanglement analysis.  The
$\lambda$ values show that the pair of filled $\sigma$ bonding and anti-bonding
orbitals do not fully localise the associated electrons.  A dashed line
indicates that the two orbitals are paired, and the $\lambda$ value is inserted
between them to indicate the degree to which localisation occurs.
(c)~Entanglement spectra of the perfect bonding theory, in which all valence
electrons are fully delocalised.  Indicated above each bar is the number of
eigenvalues in each block-diagonal sector of the spatial density matrix for
these model wave functions.  The vertical spread of each bar is proportional to
the corresponding degeneracy.  The sectors are labelled by the total number and
spin polarisation of electrons ($N$, $\text{S}_{z}$), where $N$ is equal to the number of electrons.  
For real systems core orbitals and other localized orbitals might be present. These add
extra electrons in region A and offset $N$ by a constant.
  The magnitude of the eigenvalues are all equal with values, 0.25, 0.0625, and 0.0156 for a
single bond, double bond, and triple bond respectively.  
  \label{fig:simplemodel}}
\end{figure}

The set of values for $\lambda$ provide a complete description of the bonding
fluctuations in HF theory, as the full HF wave function can be written in terms
of these modes as shown in Eq.~\ref{eq:hfent}.  We have discussed MO theory as
an incomplete theory of electron fluctuations, which can be fixed by
incorporating the idea of partial localisation.  To add this idea into standard
MO theory, we suggest simple modifications to hybridisation diagrams.  The
standard hybridisation from MO theory is shown in Fig.~\ref{fig:simplemodel}a,
and our modified diagram is shown in Fig.~\ref{fig:simplemodel}b.   Orbitals
that correspond to the standard 0, 1/2, and 1 are unchanged, as those are
already included in MO theory.  For non-standard $\lambda$ values, we indicate the two orbitals, which form the grouping, by a dashed line, and we include the $\lambda$ value
near one of the MOs.  This analysis also suggests a correspondence between the
Mayer index and Eq.~\ref{eq:hfent}, as this index also has an interpretation
based on the second moment of the particle fluctuations in the case of HF wave
functions\cite{mayer-fluc}.  From Eq.~\ref{eq:hfent} we see the entire power
spectrum of fluctuations, including the second moment, can be calculated from
the $\lambda$ values.

To analyse correlated wave functions, we begin 
by introducing model state entanglement spectra in which to compare our
numerical spectra\cite{op-2}.  The entanglement spectra are defined as the
eigenvalues of $\hat{\rho}_{A}$, and generally they are broken up into block-diagonal
sectors, which for molecules will consist of the number of electrons in region A
and the spin polarisation ($N_A, S_{z}$).  
In our entanglement theory, perfect bonds of bond order $\nu$ are given by
setting $\lambda = 1/2$ for every mode as
\begin{eqnarray}
  |\Psi^{(\nu)}\rangle = \prod_{\sigma} \prod_{i=1}^{\nu} (\sqrt{1/2}c^{\dag}_{\sigma i,A}
  +\sqrt{1/2}c^{\dag}_{\sigma i,B}) \left| 0 \right\rangle \; .
\end{eqnarray}
The core electrons and filled bonding/anti-bonding modes are described by
$\lambda=(0,0,1,1)$, and can be included into the above wave function forms
without changing the coefficients in front of the terms.
The defining characteristic of these entanglement spectra is the number of
eigenvalues in each block-diagonal sector and a single number, which is the
magnitude of these eigenvalues.  This is plotted for single, double, and triple
bonds in Fig.~\ref{fig:simplemodel}c.

\begin{figure}
  \centering
  \includegraphics[width=0.9\textwidth]{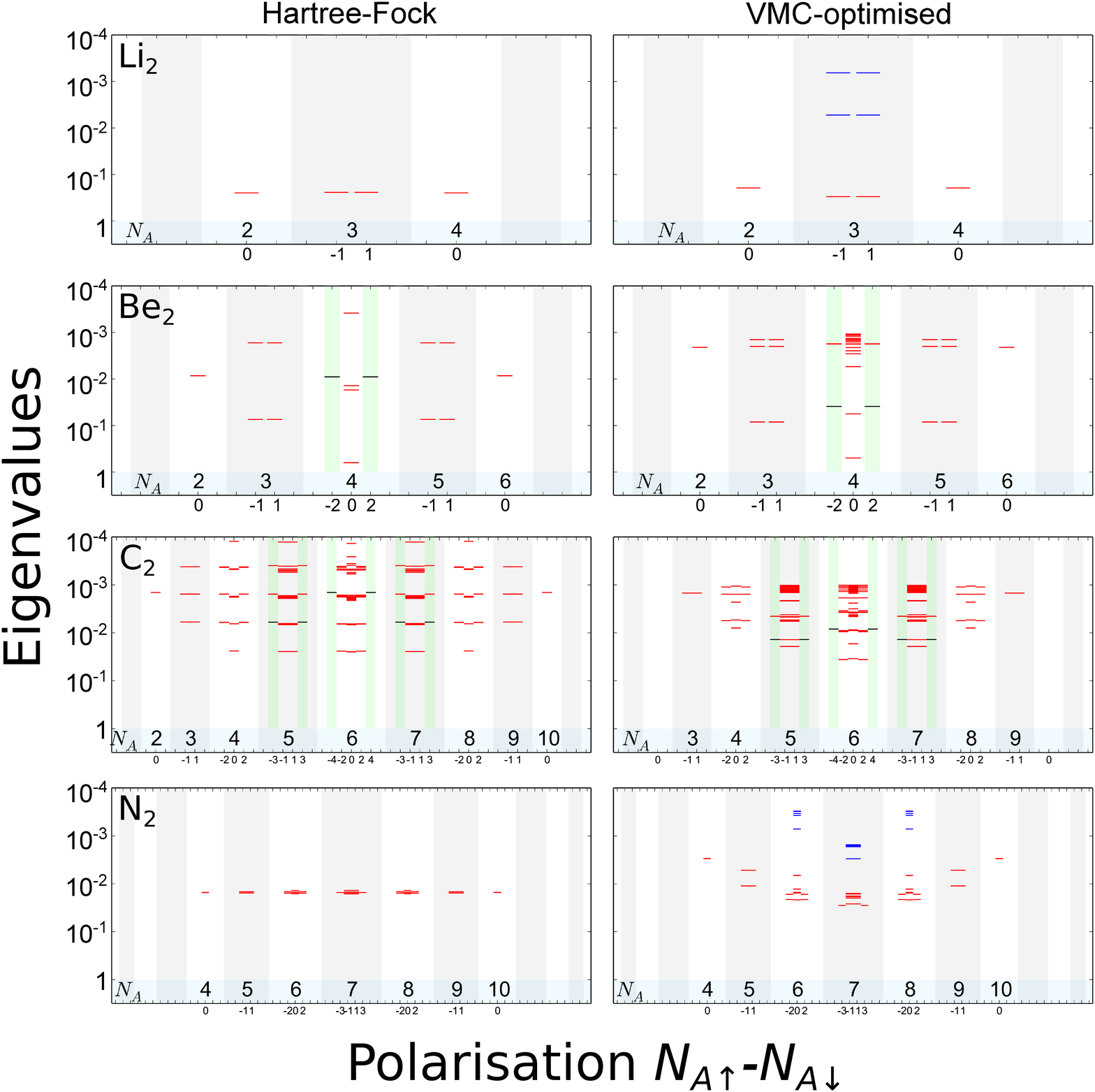} 
  \caption{Entanglement spectra of $\text{Li}_2$, $\text{Be}_2$, $\text{C}_2$,
and $\text{N}_2$.  The left and right columns are based on the HF and
VMC optimized wave functions, respectively.  Errorbars on the eigenvalues are
estimated to be no larger than $3 \times 10^{-4}$ for the VMC optimized wave functions. The magnitudes
of the eigenvalues are plotted on an inverted log scale.  The larger an eigenvalue the more important it is.  The HF wave functions
for $\text{Li}_2$ and $\text{N}_2$ have entanglement spectra that correspond to
perfect single and triple bonds, respectively.   A gap opens up in each of their
many-body entanglement spectra.  The states above and below the gap are colored blue and red respectively.  The $\text{Be}_2$ and $\text{C}_2$ molecules are more
complicated due to the partial bonds.  In comparing to the HF,  the
VMC optimized entanglement spectrum has an enhancement of anti-ferromagnetic
fluctuations in both of these molecules.  The anti-ferromagnetic sectors are
highlighted in green, and the most dominant eigenvalues in these sectors are colored black.  The magnitude of these eigenvalues are given in Table~\ref{tab:bigeig}.  
  \label{fig:entspec}}
\end{figure}

In Fig.~\ref{fig:entspec} we present the entanglement spectra of four different
molecules from two types of wave functions: the HF wave function and a wave
function optimised with VMC\cite{umrigar1996,tubman4}.  
spectra are plotted in Fig.~\ref{fig:entspec}.  A summary of different wave
functions and their energies for $\text{C}_{2}$ is given in Table~\ref{tab:ene}.
We see the HF spectrum, which can be derived from Table~\ref{tab:lambda} and
Eq.~\ref{eq:hfent}, has the same form as a perfect single bond in the case of
$\text{Li}_{2}$ and a perfect triple bond in the case of $\text{N}_{2}$.  This
picture remains essentially unchanged for the many-body wave functions.  The
effect of Coulomb repulsion reduces the probability of sectors with large charge
fluctuations.  One distinct feature present is a gap in the spectrum.  The
number of modes in each sector below the gap corresponds to that of our model
spectra.  This is similar to what is seen in the fractional quantum Hall wave
functions\cite{op-2}, in which the entanglement spectrum below the gap is called
the \emph{universal} part of the spectrum and one can count the modes in the
various sectors to determine what type of fractional quantum Hall system is
present.  The modes below the gap have the highest probability and capture the
essential physics of the bond.  This also makes this analysis useful for more
complicated molecules, which are not symmetric, as the number of eigenvalues
below the gap will remain unchanged with regards to asymmetries.  Using the
counting of the eigenmodes below the gap, we can identify $\text{N}_{2}$ as a
triple bond and $\text{Li}_{2}$ as a single bond, even for the VMC-optimised
wave functions.  Such simplicity is not evident for $\text{C}_{2}$.

There are many interesting properties of the $\text{C}_{2}$ molecule that can be
determined from the entanglement spectrum.  To address recent VB studies, we are
concerned specifically with the possibilities of anti-ferromagnetic
correlations\cite{xu:2014,epfl}, which are described as an increase in
probability of electrons of like spin accumulating on the first atom, and
likewise for the second atom but with the opposite spin.
These anti-ferromagnetic correlations can be identified with the quantum numbers
of the block diagonal sectors in $\hat{\rho}_{A}$ and we can quantitatively
compare the difference between HF, full valance CAS, and the VMC-optimised wave
functions.  From Fig.~\ref{fig:entspec} and Table~\ref{tab:bigeig} we see that
there is a significant increase in anti-ferromagnetic correlations, and in
particular the probability of the dominant eigenvalues in the sectors $N_{A} = 6$,
$S_{z} = \pm 4$ increase by an order of magnitude over the same eigenvalues in
the HF wave function.  The sectors $N_{A} = 5$, $S_{z} = \pm 3$ are also
enhanced, though less significantly, and the sectors $N_{A} = 4$,
$S_{z} = \pm 2$ show no enhancement at all.  The full-valence CAS and
VMC-optimised wave functions balance competing effects that try to reduce charge
fluctuations and enhance the anti-ferromagnetic correlations.  The previously
predicted anti-ferromagnetic correlations\cite{xu:2014} are qualitatively
correct, but the entanglement spectrum provides a detailed quantitative picture.  
The eigenvalues of the largest charge fluctuations for the VMC-optimised wave
function are severely reduced in comparison to the HF wave function, and unlike
in the case of HF, it is open to interpretation as to whether this is a triple
or quadruple bond.
Regardless, the entanglement spectrum provides useful criteria for classifying
groups of molecules, and as an example we consider the corresponding properties
between the $\text{C}_{2}$ molecule and the $\text{Be}_{2}$ molecule.  Both of
these molecules have partial bonds in their HF wave functions as seen in
Table~\ref{tab:lambda}, two inverted bonds as seen in Fig.~\ref{fig:halforbs},
and anti-ferromagnetic behavior enhanced by many-body effects as seen in the
entanglement spectrum of Fig.~\ref{fig:entspec} and Table~\ref{tab:bigeig}.
This suggests we can use the $\text{Be}_{2}$ molecule as a model for C$_{2}$ and
possibly other more complicated systems in which we want to study the behavior
of partial bonds. 

To summarise, we have developed a new way to analyse covalent bonding in
molecular systems.  For HF wave functions, this reduces to a theory of electron
fluctuations, but with a framework that allows for more complicated situations
than MO theory, such as the inclusion of partial bonds.  This analysis will be
important to understand the differences between mean field theory simulations
and many body simulations, especially for molecules that were previously
difficult to characterise.  We generalised our analysis for correlated
wave functions, and identify a gapped structure of which the eigenvalues below
the gap can be compared against model wave function entanglement spectra.  We
have provided detailed results of inverted bonds and anti-ferromagnetic
correlations in the $\text{C}_{2}$ molecule, and our tools also allowed us to
demonstrate that these bonds in $\text{C}_{2}$ observe the same physical
properties as the bonds in $\text{Be}_{2}$.  All of the entanglement tools
presented here are applicable to extended systems and more complicated bonding
situations, which are not highly symmetric. 
We believe the full impact of these tools on \textit{ab initio} systems will
change how we approach electronic structure analysis in chemistry, physics, and
materials science.

\begin{methods}

In the case of single-determinant wave functions, spatial density matrices can
be calculated analytically\cite{peschel-func,corr-2,tubman1}.  For wave
functions beyond HF, quantum Monte Carlo methods are currently the only
techniques for efficiently calculating the eigenvalues and eigenvectors of
$\hat{\rho}_{A}$ and $\hat{\rho}_{B}$ for general continuum wave functions in 2-
and 3-dimensional systems\cite{tubman6}.  For methods to calculate the related
Renyi entropies, there are several papers that describe techniques which have
been applied to molecules, Fermi-Liquids, and spin
models\cite{tubman1,tubman2,tubman3,tubman6,renyi-1}.

Quantum Monte Carlo simulations were done in our own code\cite{tubman4} which
can run variational Monte Carlo, diffusion Monte Carlo and release-node quantum
Monte Carlo.  Our QMC code can import wave functions from the GAMESS
package\cite{gamess-1}, which allowed us to simulate HF and full valence CAS
wave functions.  We separately implemented highly optimized VMC functions from a
previous work\cite{umrigar1996,tubman4}.  HF and full-valence CASSCF simulations
were run with the correlation-consistent polarized valence
quadruple-zeta~(cc-pVQZ) basis set\cite{basis-1}.  The entanglement natural
orbitals for both the HF and CASSCF can be determined from the single-particle
reduced density matrix~(1-RDM)\cite{tubman6}.  We first obtain the full HF
orbitals or natural orbitals by diagonalizing the 1-RDM.
We divide our quantum system into two regions, A and B, and the natural orbitals
are then projected onto a real space grid in region A, which are the matrix
elements of the 1-RDM in a position basis.  This matrix is then diagonalized, and
the resulting set of singular values are read off as $\sqrt{\lambda}$ for each
entanglement natural orbital.  
The matrix elements of the spatial density matrix can be calculated with the \textsc{Swap}
operator in variational quantum Monte Carlo.  We use the
entanglement natural orbitals to construct a basis in which to expand the
spatial density matrix.
The estimator for $\hat{\rho}_A$, explained in ref.~\citen{tubman6}, uses a
replica trick in which a $6N$-dimensional configuration space is used in order
to calculate the matrix elements of $\hat{\rho}_{A}$.
\end{methods}

\bibliographystyle{naturemag}
\bibliography{propref3}{}

\begin{addendum}
\item[Acknowledgements]
We would like to thank T.~Hughes, D.~Ceperley, J.~McMinis, H.~Changlani,
P.~Abbamonte and L.~Wagner for useful discussions.  This work was supported by
DARPA-OLE program and DOE DE-NA0001789.  Computer time was provided by XSEDE,
supported by the National Science Foundation Grant No. OCI-1053575, the Oak
Ridge Leadership Computing Facility at the Oak Ridge National Laboratory, which
is supported by the Office of Science of the U.S. Department of Energy under
Contract No. DE-AC05-00OR22725.  
\item[Contributions]
N.T.~designed the project and wrote the skeletal version of the QMC code.
C.Y.~built on the original code to calculate the entanglement spectra using
multi-configurational wave functions. N.T. and C.Y. both contributed to the
writing of the manuscript.
\item[Competing financial interests]
The authors declare no competing financial interests.
\item[Corresponding author]
Correspondence to: Norm Tubman
\end{addendum}

\spacing{1.25}
\begin{table}   
\centering
\begin{tabular}{c|c|c|c|}
\cline{2-4}
 & \multicolumn{3}{c|}{Electrons involved per bond type} \\ \hline
\multicolumn{1}{|c|}{Molecule} & \cellcolor[HTML]{EFEFEF}$\underset{\left( 0,0,1,1 \right)}{\text{Localised}}$
& \cellcolor[HTML]{EFEFEF}$\underset{\left( \lambda,\lambda,1-\lambda,1-\lambda \right)}{\text{Partially bonded}}$
& \cellcolor[HTML]{EFEFEF}$\underset{( 1/2 , 1/2 )}{\text{Delocalised}}$ \\ \hline \hline
\multicolumn{1}{|c|}{$\text{H}_2$} &  &  & \cellcolor[HTML]{CBCEFB}\twoBullets \\ \hline
\multicolumn{1}{|c|}{$\text{Li}_2$} & \cellcolor[HTML]{FFCCC9}\fourBullets &  & \cellcolor[HTML]{CBCEFB}\twoBullets \\ \hline
\multicolumn{1}{|c|}{$\text{Be}_2$} & \cellcolor[HTML]{FFCCC9}\fourBullets & \cellcolor[HTML]{9AFF99}{\color{blue}\fourBullets} (0.104) &  \\ \hline
\multicolumn{1}{|c|}{$\text{C}_2$} & \cellcolor[HTML]{FFCCC9}\fourBullets & \cellcolor[HTML]{9AFF99}{\color{blue}\fourBullets} (0.186) & \cellcolor[HTML]{CBCEFB}\twoBullets \, \twoBullets \\ \hline
\multicolumn{1}{|c|}{$\text{N}_2$} & \cellcolor[HTML]{FFCCC9}\fourBullets \, \fourBullets &  & \cellcolor[HTML]{CBCEFB}\twoBullets \, \twoBullets \, \twoBullets \\ \hline
\multicolumn{1}{|c|}{$\text{F}_2$} & \cellcolor[HTML]{FFCCC9}\fourBullets \, \fourBullets \, \fourBullets \, \fourBullets &  & \cellcolor[HTML]{CBCEFB}\twoBullets \\ \hline
\end{tabular}
\caption{Bonding in the first row elements 
described by the Hartree-Fock wave function.  The localised and the partially
bonded electrons exist in groups of four, while the fully delocalised electrons
exist in pairs.  The partially bonded quadruplets are labelled with their
corresponding values of $\lambda$.
\label{tab:lambda}}
\end{table}

\begin{table}   
\centering
\begin{tabular}{lll}
  Method &  Energy (hartrees)    \\
  \hline
  \cellcolor[HTML]{E0E0E0}HF & \cellcolor[HTML]{E0E0E0}-75.405 766\\
  VB\cite{xu:2014} & -75.594 679\\
  \cellcolor[HTML]{E0E0E0}Full-Valence CAS & \cellcolor[HTML]{E0E0E0}-75.643 166\\
  MRCI+Q\cite{xu:2014} & -75.803 136\\
  \cellcolor[HTML]{E0E0E0}VMC\cite{umrigar1996,tubman4} & \cellcolor[HTML]{E0E0E0}-75.8282(4)\\
  FN-DMC\cite{umrigar1996,tubman4} & -75.8901(7)\\
  RN-DMC\cite{tubman4} & -75.8969(1)\\
  FN-DMC(2008)\cite{art:9000215} & -75.9106(1)\\
  Estimated exact\cite{umrigar1996,tubman4} & -75.9265
\end{tabular}
\caption{Ground state energies of $\text{C}_2$ for various electronic structure
methods.  The wave functions used in this work are indicated as shaded cells.
The high quality VMC optimized wave function is more than 200~mHa lower in
energy than the wave function used in VB theory.  The most accurate
results given by fixed-node diffusion Monte Carlo (FN-DMC) and release-node
diffusion Monte Carlo (RN-DMC) are techniques in which the entanglement spectrum
estimators can be implemented in future work.
\label{tab:ene}}
\end{table}

\begin{table}   
\centering
\begin{tabular}{c||cc}
  {\Large $\text{Be}_2$} & \multicolumn{2}{c}{Eigenvalues} \\
Method & \begin{tabular}[c]{@{}c@{}} $N_A = 4 \pm 1$ \\ $S_z = \pm 1$ \end{tabular}
 & \begin{tabular}[c]{@{}c@{}} $N_A = 4$ \\ $S_z = \pm 2$\end{tabular} \\ \hline
\begin{tabular}[c]{@{}c@{}} Hartree-Fock \\ Full-Valence CAS \\ VMC \end{tabular}
 & \begin{tabular}[c]{@{}c@{}}0.07373 \\ 0.06460 \\ 0.08394\end{tabular}
 & \begin{tabular}[c]{@{}c@{}}0.00900 \\ 0.02440 \\ 0.03896\end{tabular}
\end{tabular}
\vspace{1cm}

\centering
\begin{tabular}{c||cccc}
  {\Large $\text{C}_2$} & \multicolumn{4}{c}{Eigenvalues} \\
Method & \begin{tabular}[c]{@{}c@{}} $N_A = 6 \pm 3$ \\ $S_z = \pm 1$ \end{tabular}
 & \begin{tabular}[c]{@{}c@{}} $N_A = 6 \pm 2$ \\ $S_z= \pm 2$ \end{tabular}
 & \begin{tabular}[c]{@{}c@{}} $N_A = 6 \pm 1$ \\ $S_z= \pm 3$ \end{tabular}
 & \begin{tabular}[c]{@{}c@{}} $N_A = 6$ \\ $S_z = \pm 4$ \end{tabular} \\ \hline
\begin{tabular}[c]{@{}c@{}}Hartree-Fock \\ Full-Valence CAS \\ VMC \end{tabular}
 & \begin{tabular}[c]{@{}c@{}}0.00593 \\ 0.00109 \\ 0.00148\end{tabular}
 & \begin{tabular}[c]{@{}c@{}}0.00605 \\ 0.00508 \\ 0.00557\end{tabular}
 & \begin{tabular}[c]{@{}c@{}}0.00596 \\ 0.01613 \\ 0.01382\end{tabular}
 & \begin{tabular}[c]{@{}c@{}}0.00144 \\ 0.01564 \\ 0.00832\end{tabular}
\end{tabular}
\vspace{1cm}

\caption{The entanglement spectrum dominant eigenvalues for the
  anti-ferromagnetic sectors in $\text{C}_{2}$ and $\text{Be}_{2}$ with
  different wave function types.  The errorbar on the eigenvalues for the
  full-valence CAS and VMC are estimated to be no larger than $3 \times
  10^{-4}$. The particle number in region~A and $S_{z}$ are the labels for the
  different sectors.  The other eigenvalues for each of these anti-ferromagnetic
  sectors are orders of magnitude smaller, and thus are not presented here.  For
  the anti-ferromagntic sectors that are charge-neutral sectors or close to
  charge neutral, there is a significant increase in the dominant eigenvalue
  when compared to HF wave functions.  For the sectors with large charge
  fluctuations, the eigenvalues are lessened because of Coulomb repulsion.
\label{tab:bigeig}}
\end{table}

\clearpage

\clearpage


\clearpage

\end{document}